\documentclass[letter]{aa}  

\usepackage{graphicx}
\usepackage{txfonts}
\usepackage{gensymb}

\begin{document}

   \title{The ALMA-PILS survey: First tentative detection of 3-hydroxypropenal (HOCHCHCHO) in the interstellar medium and chemical modeling of the C$_3$H$_4$O$_2$ isomers}
   \titlerunning{First tentative detection of HOCHCHCHO in the interstellar medium}

   \author{A. Coutens \inst{1}
          \and J.-C. Loison \inst{2} 
          \and A. Boulanger \inst{1}
          \and E. Caux \inst{1}
          \and H. S. P. M\"uller \inst{3}
          \and V. Wakelam \inst{4}
          \and S. Manigand \inst{5}
          \and J. K. J\o rgensen \inst{6}
          }

   \institute{
    Institut de Recherche en Astrophysique et Plan\'etologie (IRAP), Universit\'e de Toulouse, UPS, CNRS, CNES, 9 av. du Colonel
Roche, 31028 Toulouse Cedex 4, France \\
              \email{audrey.coutens@irap.omp.eu}
            \and Institut des Sciences Mol\'eculaires (ISM), CNRS, Universit\'e Bordeaux, 351 cours de la Lib\'eration, 33400 Talence, France
            \and I. Physikalisches Institut, Universit\"at zu K\"oln, Z\"ulpicher Str.77, 50937 K\"oln, Germany
           \and Laboratoire d'astrophysique de Bordeaux, Univ. Bordeaux, CNRS, B18N, all\'ee Geoffroy Saint-Hilaire, 33615 Pessac, France
           \and Laboratoire d'Etudes Spatiales et d'Instrumentation en Astrophysique (LESIA), Observatoire de Paris, Universit\'e PSL, CNRS, Sorbonne Universit\'e, Universit\'e de Paris, 5 place Jules Janssen, 92195 Meudon, France
           \and Niels Bohr Institute, University of Copenhagen, \O ster Voldgade 5--7, 1350 Copenhagen K, Denmark            
             }

   \date{Received xxx; accepted xxx}
 
  \abstract
  {Characterizing the molecular composition of solar-type protostars is useful for improving our understanding of the physico-chemical conditions under which the Sun and its planets formed. 
  In this work, we analyzed the Atacama Large Millimeter/submillimeter Array (ALMA) data of the Protostellar Interferometric Line Survey (PILS), an unbiased spectral survey of the solar-type protostar IRAS~16293--2422, and we tentatively detected 3-hydroxypropenal (HOCHCHCHO) for the first time in the interstellar medium towards source B. Based on the observed line intensities and assuming local thermodynamic equilibrium, its column density is constrained to be $\sim$10$^{15}$ cm$^{-2}$, corresponding to an abundance of 10$^{-4}$ relative to methanol, CH$_3$OH. Additional spectroscopic studies are needed to constrain the excitation temperature of this molecule. 
We included HOCHCHCHO and five of its isomers in the chemical network presented in \citet{Manigand2021} and we predicted their chemical evolution with the Nautilus code. The model reproduces the abundance of HOCHCHCHO within the uncertainties. This species is mainly formed through the grain surface reaction CH$_2$CHO + HCO $\rightarrow$ HCOCH$_2$CHO, followed by the tautomerization of HCOCH$_2$CHO into HOCHCHCHO. Two isomers, CH$_3$COCHO and CH$_2$COHCHO, are predicted to be even more abundant than HOCHCHCHO. Spectroscopic studies of these molecules are essential in searching for them in IRAS~16293--2422 and other astrophysical sources.}


   \maketitle
%

\section{Introduction}

Characterizing the inventory of molecules in solar-type protostars is essential for gaining a better understanding of our origins, as such studies provide important clues about the chemical richness that was available at the time when the Sun and its planets formed. Large spectral surveys are ideal for identifying the variety of molecules present in these objects, which can lead to new molecular detections in space. Indeed, the number of first detections of molecules in the interstellar medium (ISM) continues to grow thanks to the improving capacities of the instruments in the centimeter, millimeter, and submillimeter range \citep{McGuire2021}. 

\object{IRAS 16293-2422} (hereafter IRAS16293) is one of the most frequently studied solar-type protostars. The rich chemistry of this Class 0 protostellar system was studied in an unbiased way thanks to an array of large spectral surveys  \citep[e.g., IRAM-30m, JCMT, Herschel/HIFI,][]{Caux2011,Ceccarelli2010}. The Protostellar Interferometric Line Survey (PILS, \citealt{Jorgensen2016}), carried out with the Atacama Large Millimeter/submillimeter Array (ALMA), has contributed significantly to the characterization of the chemical content in this source. It revealed, for the first time, the presence of various complex organics in solar-type protostars such as c-C$_2$H$_4$O, C$_2$H$_5$CHO, CH$_3$COCH$_3$, CH$_3$NCO, NH$_2$CN, CH$_3$NC, CH$_3$OCH$_2$OH, and t-C$_2$H$_5$OCH$_3$ \citep[e.g.,][]{Lykke2017,Ligterink2017,Coutens2018,Calcutt2018,Manigand2020}. Over the last few years, the first interstellar detections of CH$_3$Cl \citep{Fayolle2017}, HONO \citep{Coutens2019}, HOCH$_2$CN \citep{Zeng2019}, and CH$_3$C(O)CH$_2$OH \citep{Zhou2020} were also obtained with ALMA towards component B of this source. More recently, \citet{Manigand2021} investigated the chemistry of three-carbon molecules in this protostar. This study led to the additional detection of propenal (C$_2$H$_3$CHO) and propylene (C$_3$H$_6$) in IRAS16293 B, while upper limits were derived for other species (C$_3$H$_8$, HCCCHO, n-C$_3$H$_7$OH, i-C$_3$H$_7$OH, C$_3$O).

In this letter, we report the first tentative interstellar detection of another three-carbon species, 3-hydroxypropenal (HOCHCHCHO) towards IRAS16293 B using the ALMA-PILS data.
This molecule is used as a biomarker to measure the level of oxidative stress in living organisms \citep[e.g.,][]{DelRio2005}. It is produced via lipid peroxidation of polyunsaturated fatty acids \citep{Pryor1975}.
The formation of this species in space is however unknown and, to our knowledge, it has never been predicted by astrochemical models. In the laboratory, C$_3$H$_4$O$_2$ was recently found to be produced in interstellar ice analogs of carbon monoxide and water at 5 K after the irradiation of energetic electrons, however, it was not possible to discriminate the isomer through the experiment \citep{Turner2021}. 

This letter is organized as follows. We present the PILS observations in Section \ref{sect_obs}. We analyze the data in Section \ref{sect_analysis}. Finally, we present the chemical modeling of HOCHCHCHO and its isomers and we discuss the results in Section \ref{sect_discu}. 

\section{Observations and spectroscopy}
\label{sect_obs}

\begin{figure}[t!]
\begin{center}
\includegraphics[width=\hsize]{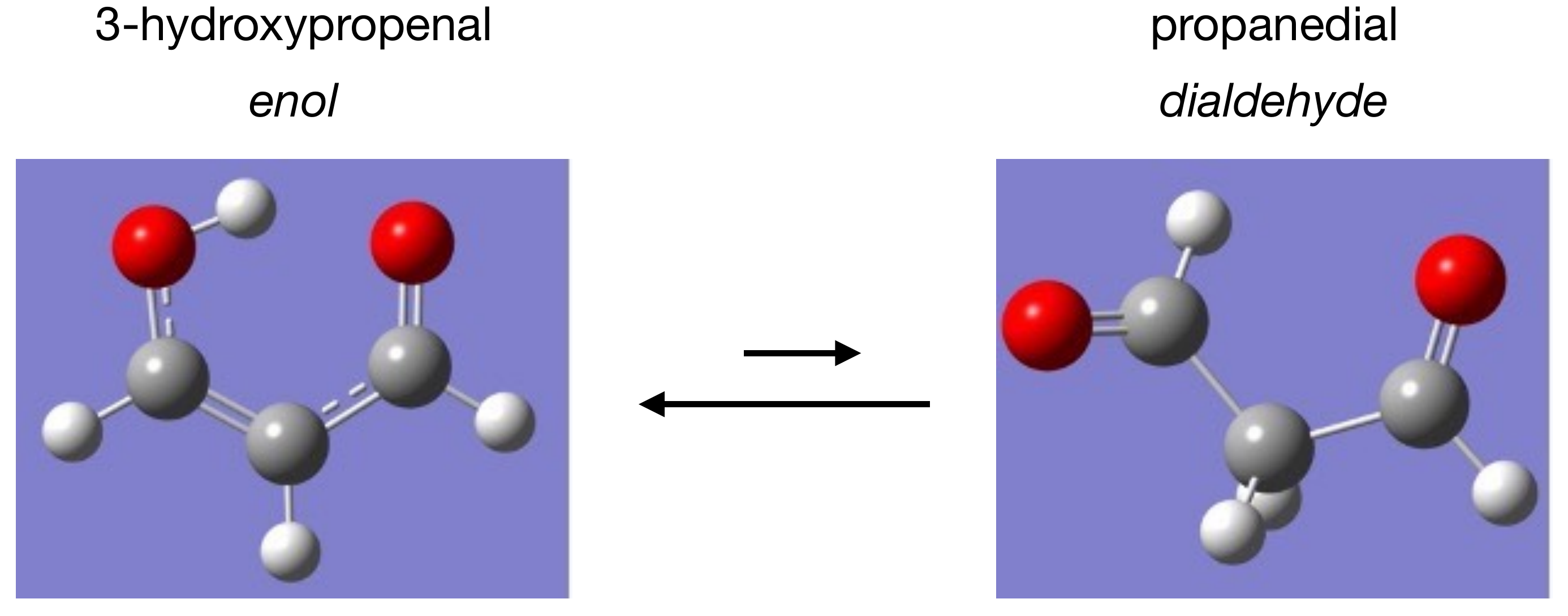}
\caption{Chemical representation of the tautomers of malonaldehyde. The enol form, 3-hydroxypropenal, is on the left, while the dialdehyde form, propanedial, is on the right.}
\label{fig_mol}
\end{center}
\end{figure}

The three-carbon species 3-hydroxypropenal is the enol tautomeric form of malonaldehyde. 
The enolic \textit{cis} form is strongly favored over the dialdehyde 
form of malonaldehyde because of the ability to form an intramolecular 
hydrogen bond (see Fig. \ref{fig_mol}). The hydrogen atom of this bond can move fairly easily 
from the CH$_2$ group to the CHO group, thus exchanging the positions 
of the two groups. 
The tunneling is only slightly hindered, which leads 
to a splitting of 647~GHz ($\approx$ 21.6~cm$^{-1}$ $\approx$ 31~K). 
The spectroscopic data of HOCHCHCHO come from the Cologne Database for Molecular Spectroscopy (CDMS, \citealt{Muller2001,Muller2005}). 
More information is available in Appendix \ref{app_spectro}.

The ALMA/PILS survey is fully described in \citet{Jorgensen2016}. In summary, these observations obtained with both the 12m array and the Atacama Compact Array (ACA) cover the entire range between 329.1 and 362.9 GHz (Band 7) with a resolution of 0.244 MHz ($\sim$0.2 km s$^{-1}$). 
The pointing center of the observations are taken to be in between IRAS16293A and IRAS16293B with both of them covered within the interferometric field of view. The sensitivity of the combined 12m array and ACA observations is about 4-5 mJy beam$^{-1}$ km s$^{-1}$.  
The beam sizes vary between 0.4 and 0.7$\arcsec$ and the observations were restored with a circular beam of 0.5$\arcsec$. Besides the Band 7 observations, small spectral windows were also covered in Bands 3 and 6. More information about the data reduction, the continuum subtraction, and these additional data can be found in \citet{Jorgensen2016}. 

\section{Analysis and results}
\label{sect_analysis}

Using a dedicated convolutional neural network (CNN) on a large set of multi-species synthetic spectra assuming local thermodynamic equilibrium (LTE,  \citealt{Boulanger2022}), we predicted the presence of HOCHCHCHO in the PILS data with a probability greater than 50\%. 
The applied spectrum corresponds to the component B of IRAS16293 at a position offset by 0.5$\arcsec$ (one beam) with respect to the continuum peak position in the south-west direction ($\alpha_{\rm J2000}$=16$^{\rm h}$32$^{\rm m}$22$\fs$58, $\delta_{\rm J2000}$=-24$\degr$28$\arcmin$32.8$\arcsec$). This position is regularly used in the PILS studies to search for molecules and isotopologues, as the lines are relatively narrow ($FWHM$ $\sim$ 1 km s$^{-1}$) and bright without being affected by absorption features \citep[e.g.,][]{Coutens2016,Lykke2017}.  We confirm that several lines can be identified as HOCHCHCHO with a classical analysis using the CASSIS\footnote{CASSIS has been developed by IRAP-UPS/CNRS \citep{Vastel2015}. \url{http://www.cassis.irap.omp.eu} } software. 
As the lines are broader towards IRAS16293 A, it is not possible to identify isolated transitions of 3-hydroxypropenal toward that source. 

\begin{figure*}[h!]
\begin{center}
\includegraphics[width=\hsize]{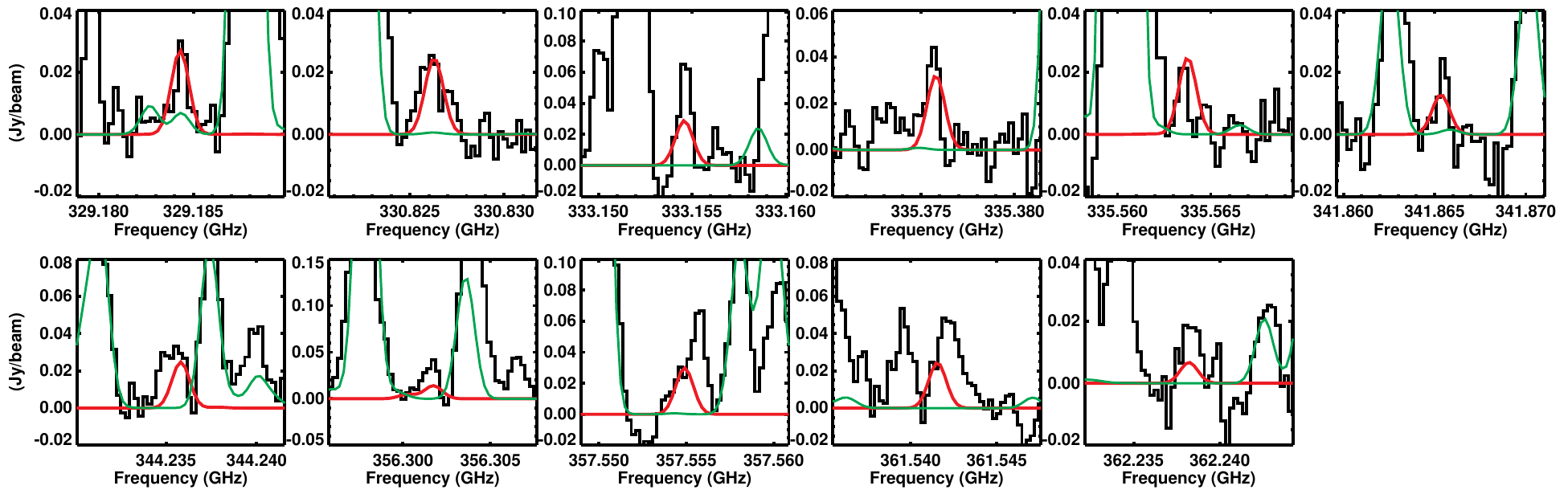}
\caption{Lines identified as HOCHCHCHO towards IRAS16293 B. The observations are in black. The model with $N$ = 1.0 $\times$ 10$^{15}$ cm$^{-2}$  and $T_{\rm ex}$ = 125 K is overlaid in red. The green line corresponds to the model which includes the previous molecules identified in the PILS survey. The lines at 333.1545, 341.8653, 356.3017, and 362.2382 GHz could be partially blended with unknown species.}
\label{fig_spectra}
\end{center}
\end{figure*}

Our analysis is similar to those presented in earlier studies \citep[e.g.,][]{Coutens2016,Coutens2019,Lykke2017,Ligterink2017}. Synthetic spectra are produced assuming LTE for various values of column densities, $N,$ and excitation temperatures, $T_{\rm ex}$. A $\chi^2$ method is then used on the unblended lines to derive the best fit parameters. To ensure that the lines are not blended with other species, we overlaid the model including the species previously identified in this source. We added HONO \citep{Coutens2019}, CH$_3$OCH$_2$OH, CH$_2$DCHO, t-C$_2$H$_5$OCH$_3$ \citep{Manigand2020}, C$_2$H$_3$CHO, C$_3$H$_6$ \citep{Manigand2021}, and CH$_3$OCHD$_2$ \citep{Richard2021} to the list presented in Appendix A of \citet{Coutens2019}. We excluded from the analysis all the transitions with large uncertainties according to the spectroscopy (see Appendix \ref{app_spectro}) those that are significantly blended with other species as well as a few that were affected by absorption features (see Table \ref{List_lines_excluded}). 
In total, we found 11 lines ($\gtrsim$ 20 mJy\,beam$^{-1}$) that could be associated to HOCHCHCHO towards IRAS16293 B (see Fig. \ref{fig_spectra}, Tables \ref{List_lines} and \ref{List_line_parameters}).

\begin{table*}[h!]
\caption{List of the transitions identified as HOCHCHCHO towards IRAS 16293 B}
\begin{center}
\begin{tabular}{cccccr}
\hline
\hline
 Transition & Frequency & Uncertainty & $E_{\rm up}$ & $A_{\rm ij}$ & $g_{\rm up}$   \\
 $N'~K_a'~K_c'~\varv_t'$ -  $N''~K_a''~K_c''~\varv_t''$ & (MHz) & (MHz) & (K) & (s$^{-1}$) & \\
 \hline
17 17 0 1 - 16 16 1 1 & 329184.2911 & 0.0223 & 171.1 & 1.30\,$\times$\,10$^{-3}$ & 35  \\
17 17 1 1 - 16 16 0 1 & 329184.2911 & 0.0223 & 171.1 & 1.29\,$\times$\,10$^{-3}$ & 105  \\
21 14 7 0 - 20 13 8 0 & 330826.2042 & 0.0188 & 148.7 & 8.32\,$\times$\,10$^{-4}$ & 129  \\
21 14 8 0 - 20 13 7 0 & 330826.5149 & 0.0188 & 148.7 & 8.32\,$\times$\,10$^{-4}$ & 43  \\
20 15 5 0 - 19 14 6 0$\dagger$ & 333154.5692 & 0.0155 & 147.3 & 9.76\,$\times$\,10$^{-4}$ & 41 \\
20 15 6 0 - 19 14 5 0$\dagger$ & 333154.5726 & 0.0155 & 147.3 & 9.76\,$\times$\,10$^{-4}$ & 123 \\
19 16 3 0 - 18 15 4 0 & 335375.7805 & 0.0154 & 146.9 & 1.14\,$\times$\,10$^{-3}$ & 117 \\
19 16 4 0 - 18 15 3 0 & 335375.7805 & 0.0154 & 146.9 & 1.14\,$\times$\,10$^{-3}$ & 39  \\
19 16 3 1 - 18 15 4 1 & 335563.7519 & 0.0154 & 177.9 & 1.12\,$\times$\,10$^{-3}$ & 39 \\
19 16 4 1 - 18 15 3 1 & 335563.7519 & 0.0154 & 177.9 & 1.12\,$\times$\,10$^{-3}$ & 117 \\
25 12 13 0 - 24 11 14 0$\dagger$ & 341865.3358 & 0.0411 & 176.0 & 5.78\,$\times$\,10$^{-4}$ & 153 \\
20 16 4 1 - 19 15 5 1 & 344235.8096 & 0.0157 & 186.3 & 1.14\,$\times$\,10$^{-3}$ & 123 \\
20 16 5 1 - 19 15 4 1 & 344235.8096 & 0.0157 & 186.3 & 1.14\,$\times$\,10$^{-3}$ & 41  \\
24 14 10 1 - 23 13 11 1$\dagger$ & 356301.7349 & 0.0304 & 208.6 & 8.58\,$\times$\,10$^{-4}$ & 147 \\ 
19 18 1 1 - 18 17 2 1 & 357554.8901 & 0.0231 & 195.8 & 1.56\,$\times$\,10$^{-3}$ & 39 \\
19 18 2 1 - 18 17 1 1 & 357554.8901 & 0.0231 & 195.8 & 1.56\,$\times$\,10$^{-3}$ & 117  \\
22 16 7 1 - 21 15 6 1 & 361541.5437 & 0.0191 & 204.3 & 1.18\,$\times$\,10$^{-3}$ & 45  \\
22 16 6 1 - 21 15 7 1 & 361541.5438 & 0.0191 & 204.3 & 1.18\,$\times$\,10$^{-3}$ & 135  \\
29 12 18 1 - 28 11 17 1$\dagger$ & 362238.2039& 0.0768 & 253.7 & 4.87\,$\times$\,10$^{-4}$ & 177  \\
\hline
\end{tabular}
\end{center}
Note: $\dagger$The best-fit model only reproduces half of the line fluxes. These lines could consequently be partially blended with unknown species. 
\label{List_lines}
\end{table*}%

To constrain the best-fit parameters, we ran two grids: one with large steps followed by another one around the best model with smaller steps ($\Delta N$ = 2 $\times$ 10$^{14}$ cm$^{-2}$, $\Delta T$ = 25 K). For this second grid, the column density ranged from 2 $\times$ 10$^{14}$ to 3 $\times$ 10$^{15}$ cm$^{-2}$, while the excitation temperature was between 50 and 350 K. To avoid overestimating the column density, we also included five undetected lines presenting small uncertainties in the $\chi^2$ analysis (see Table \ref{List_lines_other}). 
The best-fit model was obtained for a column density $N$ of 2.2 $\times$ 10$^{15}$ cm$^{-2}$ and an excitation temperature $T_{\rm ex}$ of 350 K ($\chi^2_{\rm red}$ $\sim$ 2.0). However, the excitation temperature was not well constrained. Our $\chi^2$ analysis shows that various models with excitation temperatures in the 75--350 K range can properly reproduce the lines included in the calculation (3$\sigma$ uncertainty).
It is due to the $E_{\rm up}$ values of the detected lines that only span a small range (see Table \ref{List_lines} and Table \ref{List_lines_other}). Indeed, transitions with higher $E_{\rm up}$ values are present in the PILS range but they all show large uncertainties and are consequently excluded from the analysis.
Previous molecules identified in the PILS survey were classified in two categories according to their excitation temperatures. For some molecules, the excitation temperature is found to be $\sim$125 K while for others, it is $\sim$300 K (see \citealt{Jorgensen2018} for more details). If we assume excitation temperatures of 125 and 300 K, we obtain column densities of 3-hydroxypropenal of 1.0 $\times$ 10$^{15}$ cm$^{-2}$ ($\chi^2_{\rm red}$ $\sim$ 2.0) and 1.8 $\times$ 10$^{15}$ cm$^{-2}$ ($\chi^2_{\rm red}$ $\sim$ 2.0), respectively. 
In the second hypothesis ($\geq$ 300 K), tens of additional lines should be detected. Given their large uncertainties, it is usually easy to find lines within the uncertainties that could be reproduced by the models with high $T_{\rm ex}$. However, there are a few exceptions for the transitions at 343.4656, 349.9189, and 350.0168 GHz that do not show lines within their uncertainties, which suggests that the excitation temperature could be lower than 300 K. Additional spectroscopic measurements would be needed to definitely answer this question. The model predictions obtained for $T_{\rm ex}$ = 125 K are overlaid on the spectra of the detected lines in Fig. \ref{fig_spectra} and the undetected ones in Fig. \ref{fig_spectra_undet}. Seven lines are correctly reproduced by the model and can be identified as HOCHCHCHO. The lines at 333.1546, 341.8653, 356.3017, and 362.2382 GHz are only partially reproduced, which could mean that they are partially blended with unknown species.
We carefully checked that no HOCHCHCHO line is missing with this model in other parts of the PILS survey. In Band 6, two faint lines are detected at 247.5312 and 250.4599 GHz and are in agreement with the model. Another feature is present at 239.4605 GHz but blended with an unidentified species (see Fig. \ref{fig_lines_band6}). No line is found in Band 3.

The moment 0 maps of the transitions at 329.1843, 330.8262, and 335.3758 GHz are presented in Fig. \ref{fig_map}. The emission is compact and similar to the other COMs detected in this source (see e.g., \citealt{Lykke2017,Jorgensen2018,Calcutt2018}).

The isomers of 3-hydroxypropenal were also searched for in the PILS data. HCOCH$_2$CHO (propanedial), CH$_3$COCHO (methylglyoxal), and CH$_2$COHCHO (2-hydroxypropenal) are not in the spectroscopic databases CDMS and JPL \citep{Pickett1998}. 
Only C$_2$H$_3$OCHO (vinyl formate) and the s-cis and s-trans conformers of C$_2$H$_3$COOH (propenoic acid) are available in CDMS. C$_2$H$_3$OCHO is not detected with an upper limit of 4\,$\times$\,10$^{15}$ cm$^{-2}$ based on the undetected transition at 355.2462 GHz. 
Upper limits of 1\,$\times$\,10$^{15}$ cm$^{-2}$ and 4\,$\times$\,10$^{15}$ cm$^{-2}$ are also derived for the s-cis and s-trans conformers of C$_2$H$_3$COOH, respectively.

If 3-hydroxypropenal forms on grains, its hydrogenation could potentially lead to the saturated version, 1,3-propanediol (CH$_2$OHCH$_2$CH$_2$OH). Indeed, many saturated COMs have been detected towards IRAS16293. The lower energy conformer a'GG'g present in the CDMS database was searched for and an upper limit of 1\,$\times$\,10$^{15}$ cm$^{-2}$ was derived.

\section{Chemical modeling and discussion} 
\label{sect_discu}

To better understand the chemistry of HOCHCHCHO and its isomers, we used the Nautilus code \citep{Ruaud2016}, a three-phase gas, grain surface, and grain mantle time-dependent chemical model. We updated the chemical network described in \citet{Manigand2021} by including the reactions producing and consuming the C$_3$H$_4$O$_2$ isomers as well as some radical species linked to C$_3$H$_4$O$_2$ and not yet present in the network, for example CH$_2$CHO, CH$_2$COH, or HCCHOH (see Appendix \ref{appendix_chemistry}). For the physical model of the formation of a low-mass protostar, we considered two successive stages, similarly to \citet{Manigand2021}: i) a uniform and constant stage, corresponding to the pre-stellar phase, with a density of 5 $\times$ 10$^4$ cm$^{-3}$, a temperature of 10\,K for both gas and dust, a visual extinction of 4.5 mag, a cosmic-ray ionization rate of 1.3 $\times$ 10$^{-17}$ s$^{-1}$, a standard external UV field of 1 G$_0$, the initial abundances listed in Table 3 of \citet{Manigand2021}, and a duration of 10$^6$ years; and ii) the collapse phase described in \citet{Manigand2021}.

\begin{figure}[t!]
\begin{center}
\includegraphics[width=1\hsize]{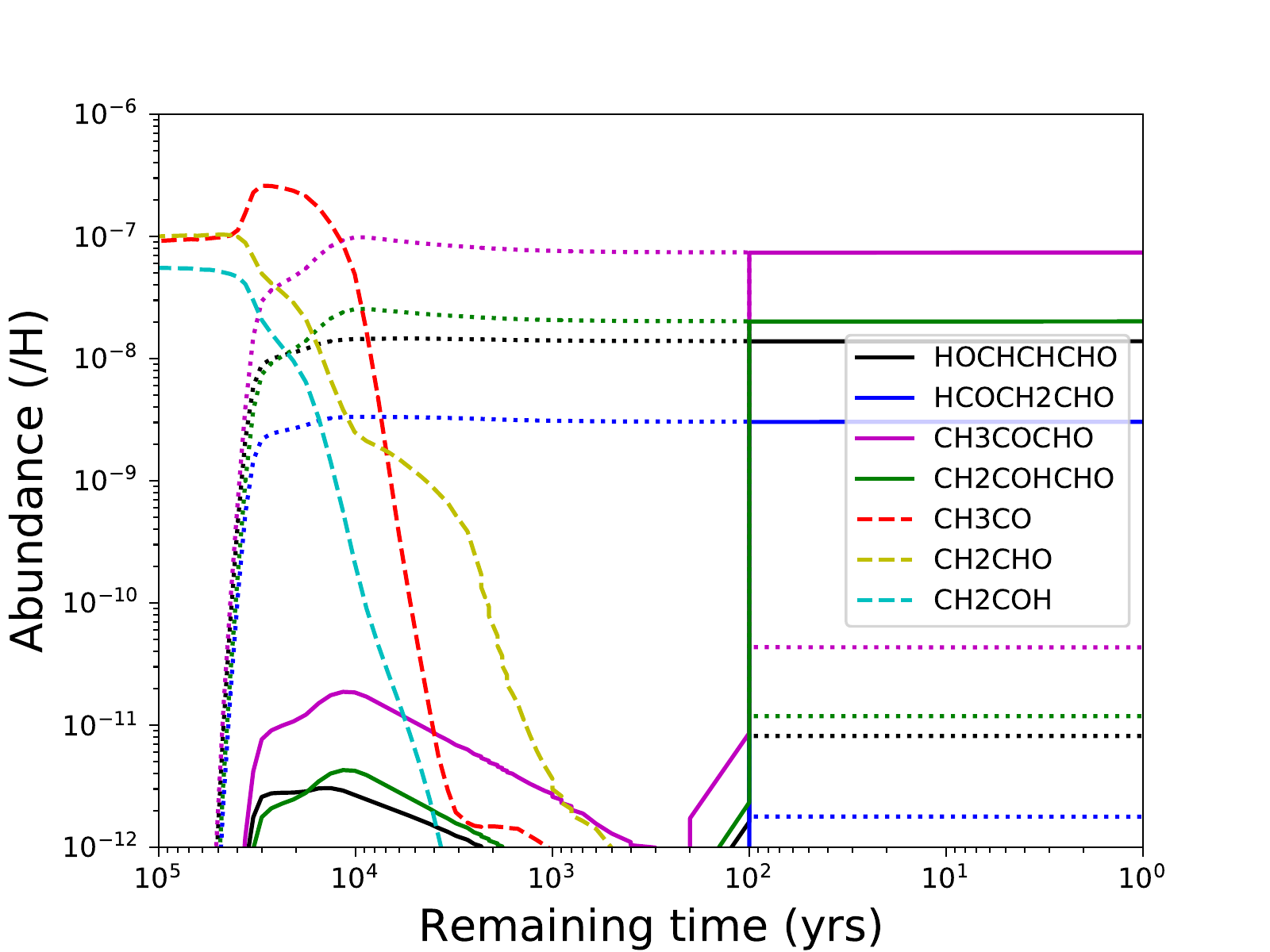}
\caption{Evolution of the abundances of the C$_3$H$_4$O$_2$ isomers, HOCHCHCHO, HCOCH$_2$CHO, CH$_3$COCHO, and CH$_2$COHCHO, and their precursors, CH$_3$CO, CH$_2$CHO, and CH$_2$COH, during the collapse phase. Solid and dotted lines are used to indicate the gas phase and grain mantle + surface abundances of the C$_3$H$_4$O$_2$ isomers, respectively. Dashed lines are used to represent the grain mantle + surface abundances of the precursors.}
\label{Fig_evol}
\end{center}
\end{figure}

Six C$_3$H$_4$O$_2$ isomers are introduced in the network: 3-hydroxypropenal (HOCHCHCHO), 2-propenoic acid (C$_2$H$_3$COOH), methyl glyoxal (CH$_3$COCHO), propanedial (HCOCH$_2$CHO), 2-hydroxypropenal (CH$_2$COHCHO), and vinyl formate (C$_2$H$_3$OCHO). The description of their formation and destruction pathways is presented in Appendix \ref{appendix_chemistry}. HOCHCHCHO is mainly formed through an indirect way through the reaction CH$_2$CHO + HCO  $\rightarrow$ HCOCH$_2$CHO followed by the tautomerization of HCOCH$_2$CHO into HOCHCHCHO. 
The predicted abundances with respect to CH$_3$OH are indicated in Table \ref{Abundances_predictions}. Among the different isomers of C$_3$H$_4$O$_2$, three are relatively abundant according to the model. HOCHCHCHO has a predicted abundance with respect to CH$_3$OH of 4.6\,$\times$\,10$^{-4}$ in agreement with the observations (1\,$\times$\,10$^{-4}$) (see below for the discussions on uncertainties). In addition, CH$_3$COCHO and CH$_2$COHCHO also show high abundances (with respect to CH$_3$OH) of 2\,$\times$\,10$^{-3}$ and 6\,$\times$\,10$^{-4}$, respectively. Future detections could be possible for these two species if the spectroscopic data are available. It should be noted that the calculated structure of propanedial (HCOCH$_2$CHO) is non-planar. Its dipole moment is very uncertain because it depends strongly on the dihedral angle between the two -CHO groups\footnote{\url{https://cccbdb.nist.gov/dipole2x.asp}}. Consequently, HCOCH$_2$CHO which is predicted to have a relatively high abundance (with respect to CH$_3$OH) of 1\,$\times$\,10$^{-4}$ may be also detectable. Figure \ref{Fig_evol} shows the evolution of the most abundant isomers and their precursors during the collapse phase. 

\begin{table*}[h!]
\caption{Predicted and observed abundances of the C$_3$H$_4$O$_2$ isomers.}
\begin{center}
\begin{tabular}{lllll}
\hline
\hline
Isomer & Formula & Predicted abundance & Predicted abundance  & Observed abundance \\
& & (/H) & (/CH$_3$OH) &  (/CH$_3$OH) \\
\hline
3-Hydroxypropenal & HOCHCHCHO      & 1.4 $\times$ 10$^{-8}$  &  4.6 $\times$ 10$^{-4}$    & 1.0 $\times$ 10$^{-4}$  \\
Propanedial & HCOCH$_2$CHO            & 3.0 $\times$ 10$^{-9}$ &  9.9 $\times$ 10$^{-5}$  &  \\
Methyl glyoxal  & CH$_3$COCHO         & 7.4 $\times$ 10$^{-8}$  &   2.4 $\times$ 10$^{-3}$   &    \\
2-Hydroxypropenal & CH$_2$COHCHO & 2.0 $\times$ 10$^{-8}$ & 6.6 $\times$ 10$^{-4}$  \\
Vinyl formate & C$_2$H$_3$OCHO       & 8.4 $\times$ 10$^{-10}$ & 2.8 $\times$ 10$^{-5}$ & $\leq$ 4 $\times$ 10$^{-4}$  \\
2-Propenoic acid & C$_2$H$_3$COOH & 2.2 $\times$ 10$^{-13}$ & 7.3 $\times$ 10$^{-9}$ & $\leq$ 5 $\times$ 10$^{-4}$ \\
\hline
\end{tabular}
\tablefoot{ The total uncertainty on the predicted abundances is about a factor 10.}
\end{center}
\label{Abundances_predictions}
\end{table*}%

The formation pathways of CH$_3$COCHO, particularly the grain surface reaction CH$_3$CO + HCO, are credible and the calculated abundance of CH$_3$COCHO makes it a potentially detectable species. For CH$_2$COHCHO, it is more ambiguous because one of the main pathways of formation is the isomerization of CH$_3$COCHO, which could be very less favorable on grains that the result given by the methodology used here.  The abundances of C$_3$H$_4$O$_2$ isomers are linearly dependent on the branching ratios of the association reactions, which are not known but whose uncertainty can be estimated at about a factor of 2 in this case. The abundances are also highly sensitive to the reactions with hydrogen atoms, which are the most efficient destruction reactions in our model. The barriers were calculated using the DFT method associated with the M06-2X functional. As these reactions take place through tunneling, they are dependent on the barrier time and the barrier width estimated by calculating the imaginary frequency. The uncertainty on the rate of these reactions is therefore not negligible and the uncertainty induced on the C$_3$H$_4$O$_2$ abundances can be estimated to be about a factor of 5. Thus, the total uncertainty on the predicted abundances of the C$_3$H$_4$O$_2$ isomers is about a factor of 10. 

In conclusion, this study confirms that solar-type protostars can harbor a broad variety of complex organic molecules in their warm inner regions, highlighting the importance of additional spectroscopic studies of three-carbon species in the submillimeter range.

\begin{acknowledgements}
This paper makes use of the ALMA data ADS/JAO.ALMA\#2013.1.00278.S and ADS/JAO.ALMA\#2012.1.00712.S. ALMA is a partnership of ESO (representing its member states), NSF (USA) and NINS (Japan), together with NRC (Canada) and NSC and ASIAA (Taiwan), in cooperation with the Republic of Chile. The Joint ALMA Observatory is operated by ESO, AUI/NRAO, and NAOJ.
This project has received funding from the European Research Council (ERC) under the European Union's Horizon 2020 research and innovation programme (grant agreement 949278, Chemtrip). VW acknowledges the CNRS program "Physique et Chimie du Milieu Interstellaire" (PCMI) co-funded by the Centre National d'Etudes Spatiales (CNES). JKJ acknowledges support from the Independent Research Fund Denmark (grant No. DFF0135-00123B).
\end{acknowledgements}

\bibliographystyle{aa} 
\bibliography{Biblio} 

\appendix

\section{Spectroscopy of 3-hydroxypropenal}
\label{app_spectro}

The spectroscopy of 3-hydroxypropenal is based on the measurements reported in \citet{Baughcum1978}, \citet{Stolze1983}, \citet{Turner1984}, \citet{Firth1991}, and \citet{Baba1999}.  The b-dipole moment component comes from \citet{Baughcum1981} while the a-dipole moment was determined by \citet{Baba1999}.  
The strong $b$-type transitions occur within the tunneling states 
(rotational transitions), whereas the weak $a$-type transitions connect 
the two tunneling states (rotation-tunneling transitions). 
The extensive set of experimental rotational transitions cover a 
large range of $J$ quantum numbers; however, transitions with 
the lowest $K_a$ were only accessed for low values of $J$ as these 
are somewhat weaker than higher-$K_a$ transitions in the same range. 
In addition, the small amount of rotation-tunneling transitions 
have also high values of $K_a$. Therefore, transitions having 
high values of $J$ and very low values of $K_a$ have large 
uncertainties which prohibit their unambiguous identification 
in the ALMA/PILS data sets currently, even though their intensities 
should be sufficiently above the noise limit in at least some cases. 

\section{Additional tables and figures}

\begin{figure*}[h!]
\begin{center}
\includegraphics[width=0.8\hsize]{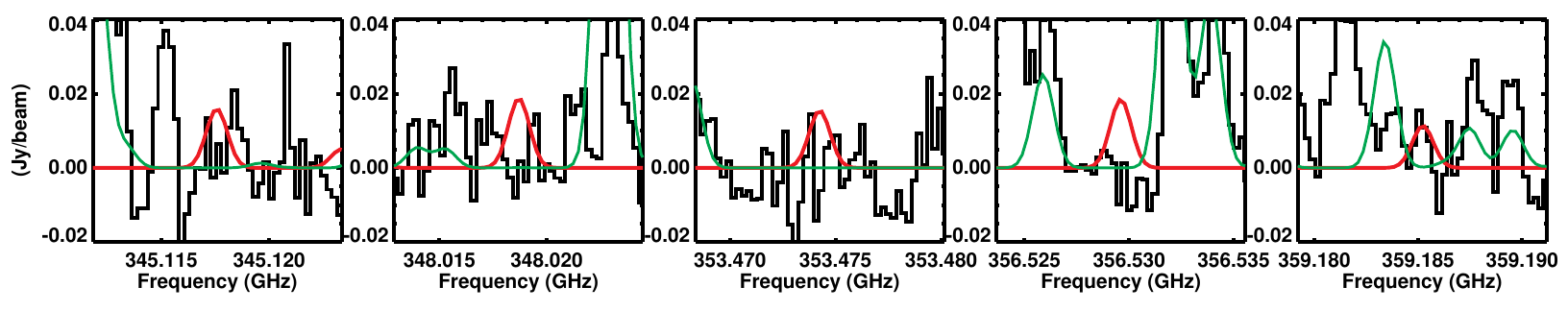}
\caption{Observations of the undetected HOCHCHCHO transitions used in the $\chi^2$ analysis (black). The model with $N$ = 1.0 $\times$ 10$^{15}$ cm$^{-2}$  and $T_{\rm ex}$ = 125 K is overlaid in red. The green line corresponds to the model including the previous molecules identified in the PILS survey.}
\label{fig_spectra_undet}
\end{center}
\end{figure*}

\begin{figure*}[h!]
\begin{center}
\includegraphics[width=0.52\hsize]{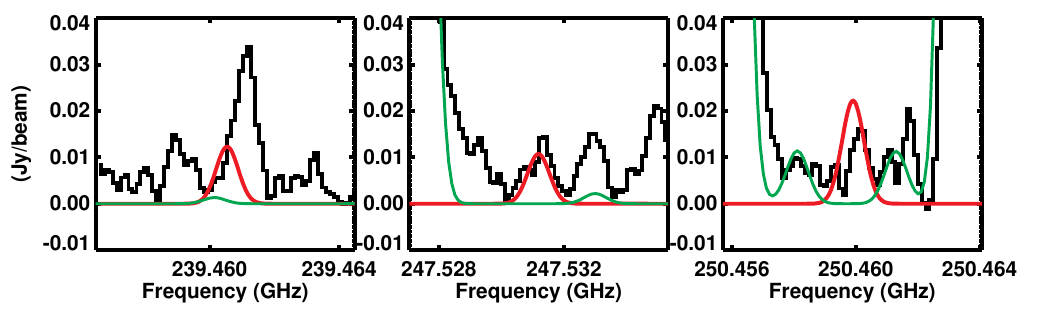}
\caption{Lines of HOCHCHCHO identified in Band 6 (black). The model with $N$ = 1.0 $\times$ 10$^{15}$ cm$^{-2}$  and $T_{\rm ex}$ = 125 K is overlaid in red. The green line corresponds to the model including the previous molecules identified in the PILS survey.}
\label{fig_lines_band6}
\end{center}
\end{figure*}

\begin{figure*}[h!]
\begin{center}
\includegraphics[width=0.95\hsize]{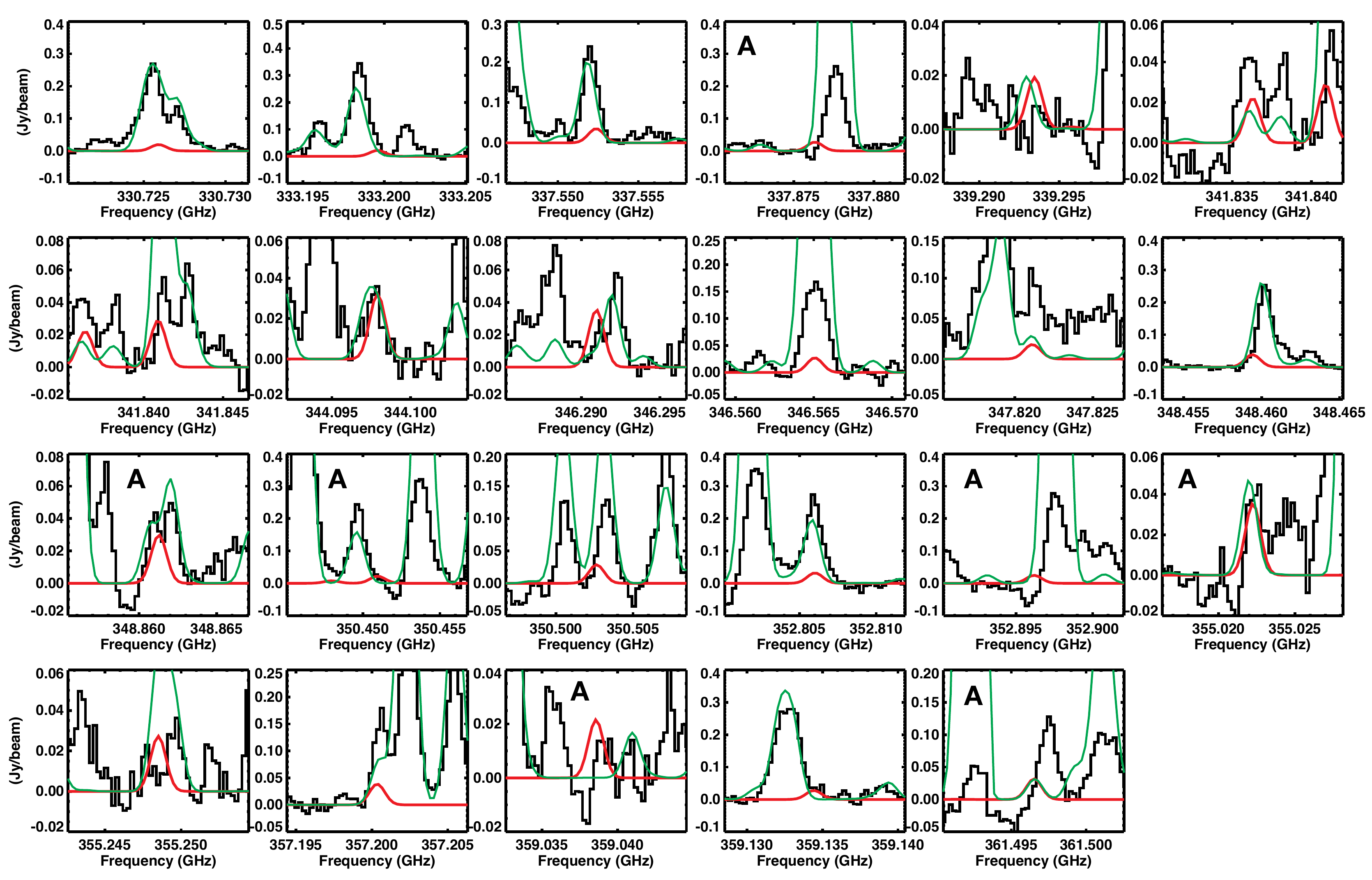}
\caption{Observations of the other HOCHCHCHO transitions that are blended and/or affected by absorption features (black). The letter A is added on the panel where significant absorption features are seen at the half-beam offset position. The model with $N$ = 1.0 $\times$ 10$^{15}$ cm$^{-2}$  and $T_{\rm ex}$ = 125 K is overlaid in red. The green line corresponds to the model including the previous molecules identified in the PILS survey. }
\label{fig_lines_band7_other}
\end{center}
\end{figure*}

\begin{figure}[!ht] 
 \begin{center} 
\includegraphics[width=\hsize]{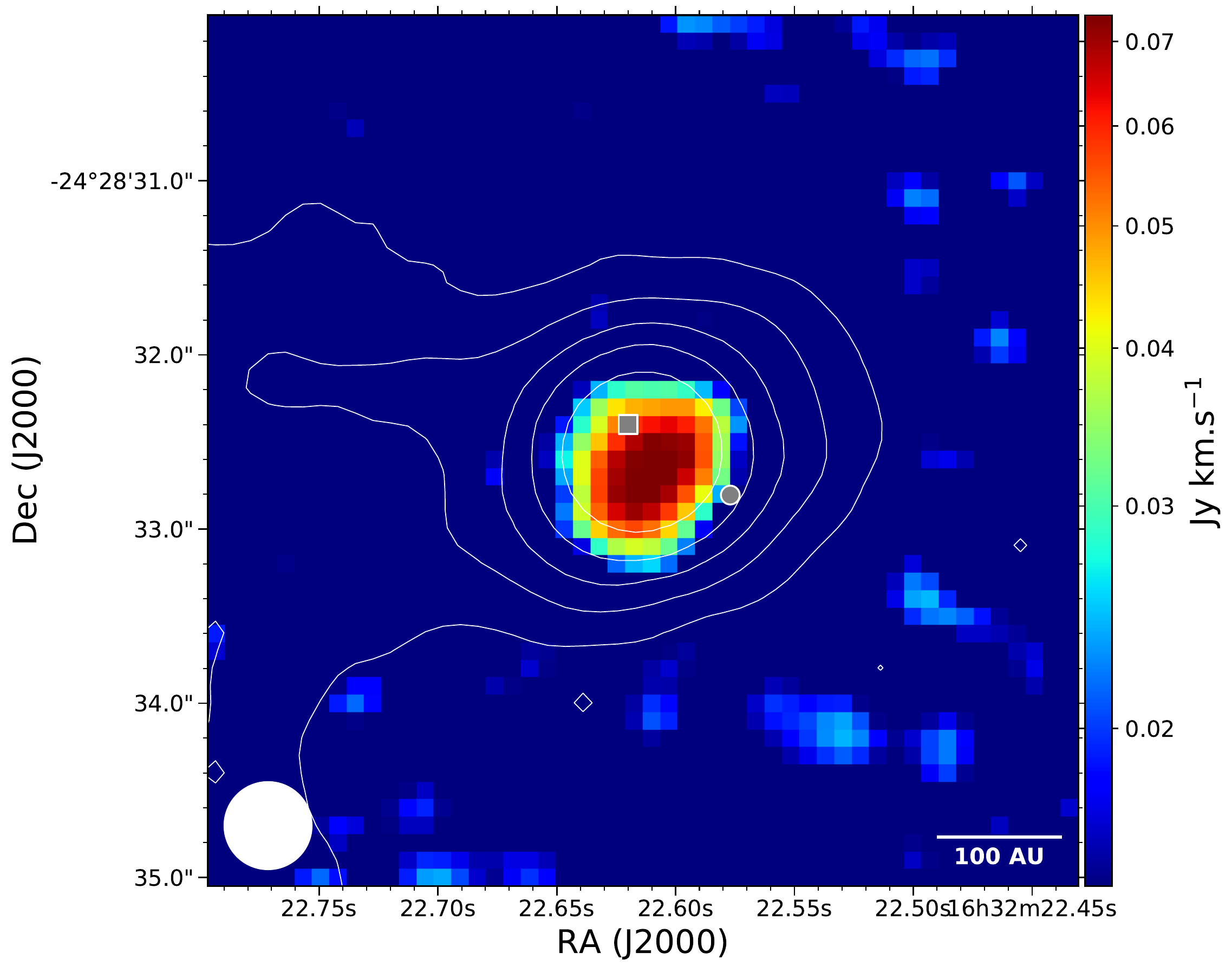}
\includegraphics[width=\hsize]{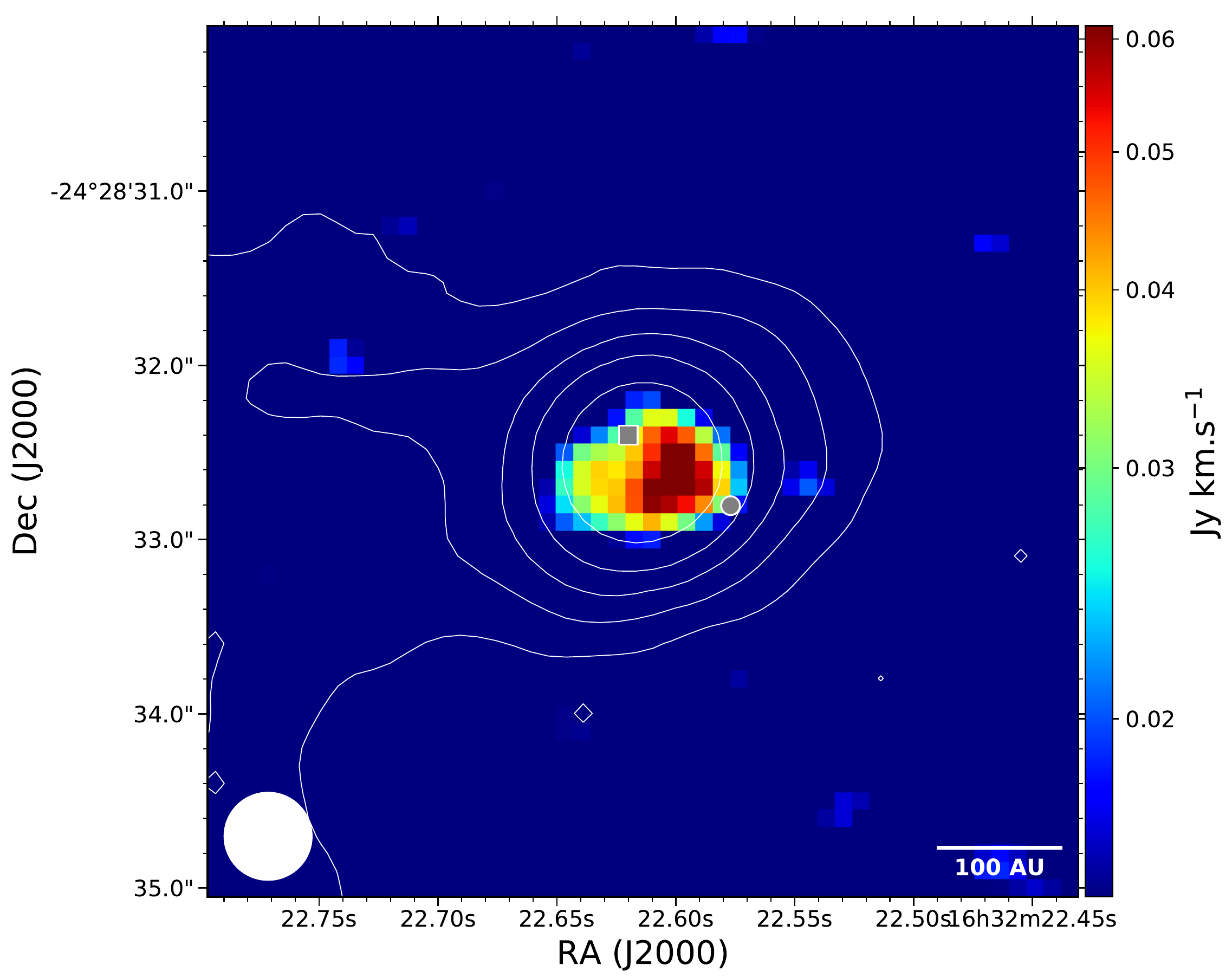}
\includegraphics[width=\hsize]{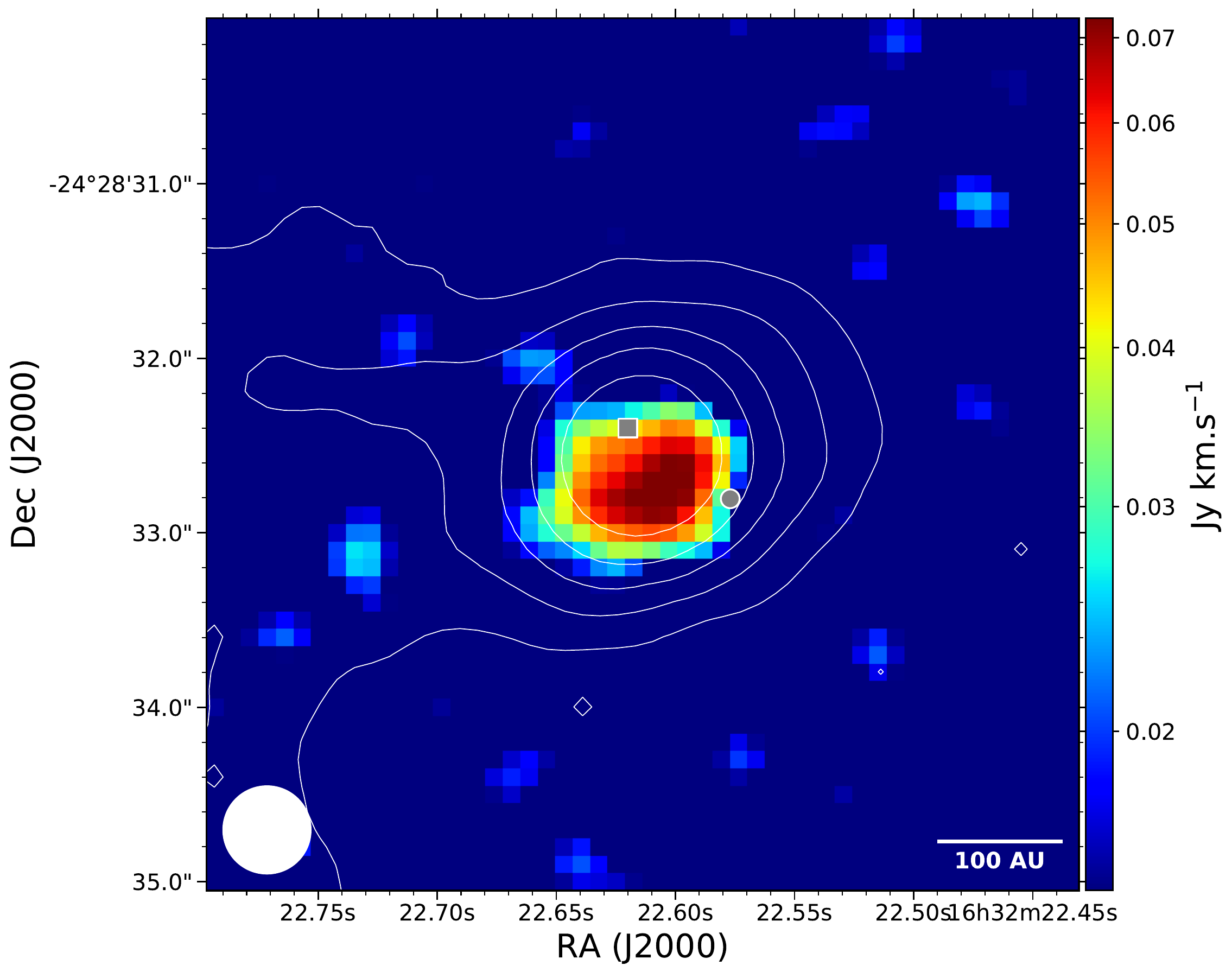}
 \caption{Integrated intensity maps of the HOCHCHCHO lines at 329.1843 (top), 330.8262 (middle), and 335.3758 GHz (bottom) towards IRAS16293 B. Dust continuum contours are indicated with white contours (levels of 0.02, 0.05, 0.1, 0.2, and 0.5 Jy beam$^{-1}$). The position of the continuum peak is indicated with a gray square, while the position analyzed for IRAS16293 B (full-beam offset) is indicated with a gray circle. The beam size is indicated in white in the bottom left corner.}
 \label{fig_map}
  \end{center} 
 \end{figure}

Table \ref{List_line_parameters} lists the observational parameters of the detected HOCHCHCHO lines obtained after Gaussian fitting. 
The undetected transitions of HOCHCHCHO used in the calculations presented in Section \ref{sect_analysis} (see Table \ref{List_lines_other}) are shown in Fig. \ref{fig_spectra_undet}. The additional lines detected in Band 6 (see Table \ref{List_lines_Band6}) are shown in Fig. \ref{fig_lines_band6}. The other transitions of HOCHCHCHO that are either blended or affected by absorption features (see Table \ref{List_lines_excluded}) are shown in Fig. \ref{fig_lines_band7_other}. 
The model that includes the previously detected molecules (in green) is known to overproduce the intensities of optically thick lines. The spectroscopy can also be uncertain in some cases. 
Maps of three HOCHCHCHO transitions are shown in Fig. \ref{fig_map}. 

\begin{table*}[h!]
\caption{Observed line parameters of the HOCHCHCHO transitions identified towards IRAS 16293 B.}
\begin{center}
\begin{tabular}{cccccc}
\hline
\hline
 Transition & Frequency &  Integrated flux & $I_{\rm max}$ & $\varv$$_{\rm LSR}$ & $FWHM$   \\
 $N'~K_a'~K_c'~\varv_t'$ -  $N''~K_a''~K_c''~\varv_t''$ & (MHz) & (Jy beam$^{-1}$ km s$^{-1}$) & (Jy beam$^{-1}$) & (km s$^{-1}$) & (km s$^{-1}$)   \\
 \hline
17 17 0 1 - 16 16 1 1 & 329184.3 &  0.024 & 0.028 $\pm$ 0.004 & 2.64 $\pm$ 0.05 & 0.72 $\pm$ 0.11 \\
17 17 1 1 - 16 16 0 1 & 329184.3 &  \\
21 14 7 0 - 20 13 8 0 & 330826.2 &  0.037 & 0.022 $\pm$ 0.003 & 2.76 $\pm$ 0.08 & 1.61 $\pm$ 0.21 \\
21 14 8 0 - 20 13 7 0 & 330826.5 &   \\
20 15 5 0 - 19 14 6 0$\dagger$ & 333154.6 &  0.045 & 0.069 $\pm$ 0.007 & 2.66 $\pm$ 0.03 & 0.63 $\pm$ 0.07 \\
20 15 6 0 - 19 14 5 0$\dagger$ & 333154.6 &  \\
19 16 3 0 - 18 15 4 0 & 335375.8 &  0.032 & 0.045 $\pm$ 0.006 & 2.88 $\pm$ 0.05 & 0.70 $\pm$ 0.11 \\
19 16 4 0 - 18 15 3 0 & 335375.8 &   \\
19 16 3 1 - 18 15 4 1 & 335563.8 & 0.016 & 0.024 $\pm$ 0.005 & 2.95 $\pm$ 0.05 & 0.51 $\pm$ 0.12 \\
19 16 4 1 - 18 15 3 1 & 335563.8 &  \\
25 12 13 0 - 24 11 14 0$\dagger$ & 341865.3  & 0.014 & 0.025 $\pm$ 0.005 & 2.60 $\pm$ 0.04 & 0.47 $\pm$ 0.10 \\
20 16 4 1 - 19 15 5 1$\ddagger$ & 344235.8 & 0.020 & 0.033 $\pm$ 0.003 & 2.81 $\pm$ 0.09 & 0.76 $\pm$ 0.19 \\
20 16 5 1 - 19 15 4 1$\ddagger$ & 344235.8 &   \\
24 14 10 1 - 23 13 11 1$\dagger$ & 356301.7 & 0.032 & 0.040 $\pm$ 0.006 & 2.87 $\pm$ 0.06 & 0.78 $\pm$ 0.14 \\ 
19 18 1 1 - 18 17 2 1$\ddagger$ & 357554.9  &  $<$ 0.067 & 0.026 $\pm$ 0.003 & 2.89 $\pm$ 0.15 & 1.10 $\pm$ 0.35 \\
19 18 2 1 - 18 17 1 1$\ddagger$ & 357554.9  &   \\
22 16 7 1 - 21 15 6 1$\ddagger$ & 361541.5 & $<$ 0.045 & 0.023 $\pm$ 0.031 & [2.7] & [0.7] \\
22 16 6 1 - 21 15 7 1$\ddagger$ & 361541.5 &   \\
29 12 18 1 - 28 11 17 1$\dagger$ & 362238.2 &  0.019 & 0.020 $\pm$ 0.003 & 2.62 $\pm$ 0.07 & 0.96 $\pm$ 0.16  \\
\\
13 12 1 1 - 12 11 2 1$\dagger$$\ddagger$ & 239460.6  & < 0.020 & 0.008 $\pm$ 0.003 & [2.7] & [0.7] \\
13 12 2 1 - 12 11 1 1$\dagger$$\ddagger$ & 239460.6  &  \\
18 9 10 0 - 17 8 9 0~~ & 247531.2  & 0.012 & 0.011 $\pm$ 0.001 & 2.59 $\pm$ 0.06 & 1.19 $\pm$ 0.17  \\
13 13 0 1 - 12 12 1 1 & 250459.9 & 0.012 & 0.015 $\pm$ 0.001 & 2.43 $\pm$ 0.05 & 0.87 $\pm$ 0.13 \\
13 13 1 1 - 12 12 0 1 & 250459.9  & \\
\hline
\end{tabular}
\end{center}
Note: The values in square brackets are assumed. The uncertainties correspond to the fit uncertainties only. $\dagger$ These lines could be partially blended with unknown species. $\ddagger$  A double Gaussian is used to fit the observations.
\label{List_line_parameters}
\end{table*}%

\begin{table*}[h!]
\caption{List of the undetected transitions of HOCHCHCHO used in the $\chi^2$ analysis.}
\begin{center}
\begin{tabular}{cccccrl}
\hline
\hline
 Transition & Frequency & Uncertainty & $E_{\rm up}$ & $A_{\rm ij}$ & $g_{\rm up}$   \\
 $N'~K_a'~K_c'~\varv_t'$ -  $N''~K_a''~K_c''~\varv_t''$ & (MHz) & (MHz) & (K) & (s$^{-1}$) & \\
 \hline
24 13 12 0 - 23 12 11 0 & 345117.5624 & 0.0325 & 171.4 & 7.24\,$\times$\,10$^{-4}$ & 147 \\
~23 14 9 0 - 22 13 10 0 & 348018.6951 & 0.0263 & 167.9 & 8.64\,$\times$\,10$^{-4}$ & 141 \\
25 13 12 0 - 24 12 13 0 & 353474.2386 & 0.0377 & 182.1 & 7.34\,$\times$\,10$^{-4}$ & 153 \\
24 14 11 0 - 23 13 10 0 & 356529.5940 & 0.0301 & 178.2 & 8.87\,$\times$\,10$^{-4}$ & 147  \\
27 12 15 0 - 26 11 16 0 & 359185.2111 & 0.0572 & 199.1 & 5.72\,$\times$\,10$^{-4}$ & 165 \\
\hline
\end{tabular}
\end{center}
\label{List_lines_other}
\end{table*}%

\begin{table*}[h!]
\caption{List of the transitions identified as HOCHCHCHO in Band 6.}
\begin{center}
\begin{tabular}{cccccrl}
\hline
\hline
 Transition & Frequency & Uncertainty & $E_{\rm up}$ & $A_{\rm ij}$ & $g_{\rm up}$   \\
 $N'~K_a'~K_c'~\varv_t'$ -  $N''~K_a''~K_c''~\varv_t''$ & (MHz) & (MHz) & (K) & (s$^{-1}$) & \\
 \hline
13 12 1 1 - 12 11 2 1 & 239460.5534 & 0.0102 & 107.0 & 4.43\,$\times$\,10$^{-4}$ & 27 \\
13 12 2 1 - 12 11 1 1 & 239460.5534 & 0.0102 & 107.0 & 4.43\,$\times$\,10$^{-4}$ & 81 \\
18 9 10 0 - 17 8 9 0 & 247531.1948 & 0.0167 & ~~93.9 & 2.26\,$\times$\,10$^{-4}$ & 111 \\
13 13 0 1 - 12 12 1 1 &  250459.9239 & 0.0124 & 113.6 & 5.59\,$\times$\,10$^{-4}$ & 27 \\
13 13 1 1 - 12 12 0 1 &  250459.9239 & 0.0124 & 113.6 & 5.59\,$\times$\,10$^{-4}$ & 81 \\
\hline
\end{tabular}
\end{center}
\label{List_lines_Band6}
\end{table*}%

\begin{table*}[h!]
\caption{List of the HOCHCHCHO transitions excluded from the analysis due to blending with other species or absorption features.}
\begin{center}
\begin{tabular}{cccccrl}
\hline
\hline
Transition & Frequency & Uncertainty & $E_{\rm up}$ & $A_{\rm ij}$ & $g_{\rm up}$ & Reason for exclusion \\
  & (MHz) & (MHz) & (K) & (s$^{-1}$) & & \\
 \hline
21 14 8 1 - 20 13 7 1 & 330725.9000 & 0.0187 & 179.4 & 8.11\,$\times$\,10$^{-4}$ & 129 & Blended with C$_2$H$_5$OH \\
21 14 7 1 - 20 13 8 1 & 330725.9273 & 0.0187 & 179.4 & 8.1\,$\times$\,10$^{-4}$ & 43 & and CH$_3$COCH$_3$ \\
20 15 6 1 - 19 14 5 1 & 333199.5740 & 0.0154 & 178.2 & 9.56\,$\times$\,10$^{-4}$ & 41 & Blended with CH$_2$DOCH$_3$ \\
20 15 5 1 - 19 14 6 1 & 333199.5741 & 0.0154 & 178.2 & 9.56\,$\times$\,10$^{-4}$ & 123 \\
18 17 1 0 - 17 16 2 0 & 337552.3721 & 0.0200 & 147.5 & 1.32\,$\times$\,10$^{-3}$ & 37 & Blended with D$_2^{13}$CO \\
18 17 2 0 - 17 16 1 0 & 337552.3721 & 0.0200 & 147.5 & 1.32\,$\times$\,10$^{-3}$ & 111 \\
18 17 2 1 - 17 16 1 1 & 337876.3115 & 0.0200 & 178.7 & 1.31\,$\times$\,10$^{-3}$ & 37 & Blended with CH$_3$OH \\
18 17 1 1 - 17 16 2 1 & 337876.3115 & 0.0200 & 178.7 & 1.31\,$\times$\,10$^{-3}$ & 111 & + Absorption \\
22 14 9 1 - 21 13 8 1 & 339293.3621 & 0.0219 & 188.7 & 8.27\,$\times$\,10$^{-4}$ & 45 & Blended with NHDCN \\
22 14 8 1 - 21 13 9 1 & 339293.4748 & 0.0219 & 188.7 & 8.27\,$\times$\,10$^{-4}$ & 135 \\
21 15 6 1 - 20 14 7 1 & 341836.2789 & 0.0175 & 187.0 & 9.75\,$\times$\,10$^{-4}$ & 129 & Blended with $^{13}$CH$_3$OH \\
21 15 7 1 - 20 14 6 1 & 341836.2797 & 0.0175 & 187.0 & 9.75\,$\times$\,10$^{-4}$ & 43 \\
21 15 6 0 - 20 14 7 0 & 341840.8890 & 0.0176 & 156.2 & 9.99\,$\times$\,10$^{-4}$ & 129 & Blended with CH$_3$OCHO \\
21 15 7 0 - 20 14 6 0 & 341840.9043 & 0.0176  & 156.2 & 9.99\,$\times$\,10$^{-4}$ & 43 \\
20 16 5 0 - 19 15 4 0 & 344097.8858 & 0.0157 & 155.4 & 1.16\,$\times$\,10$^{-3}$ & 123 & Blended with CH$_3$OCHO \\
20 16 4 0 - 19 15 5 0 & 344097.8858 & 0.0157 & 155.4 & 1.16\,$\times$\,10$^{-3}$ & 41 \\
19 17 2 0 - 18 16 3 0 & 346290.9455 & 0.0183 & 155.5 & 1.35\,$\times$\,10$^{-3}$ & 117 & Blended with CH$_2$DOH \\
19 17 3 0 - 18 16 2 0 & 346290.9455 & 0.0183 & 155.5 & 1.35\,$\times$\,10$^{-3}$ & 39 \\
19 17 2 1 - 18 16 3 1 & 346565.0565 & 0.0183 & 186.6 & 1.33\,$\times$\,10$^{-3}$ & 39 & Blended with C$_2$H$_5$OH \\
19 17 3 1 - 18 16 2 1 & 346565.0565 & 0.0183 & 186.6 & 1.33\,$\times$\,10$^{-3}$ & 117 & and CH$_3$OCHO \\
23 14 10 1 - 22 13 9 1 & 347821.0130 & 0.0258 & 198.4 & 8.43\,$\times$\,10$^{-4}$ & 141 & Blended with (CH$_2$OH)$_2$ \\
23 14 9 1 - 22 13 10 1 & 347821.4373 & 0.0258 & 198.4 & 8.43\,$\times$\,10$^{-4}$ & 47 \\
18 18 0 0 - 17 17 1 0 & 348459.4070 & 0.0257 & 156.7 & 1.56\,$\times$\,10$^{-3}$ & 37 & Blended with CH$_2$DCHO \\
18 18 1 0 - 17 17 0 0 & 348459.4070 & 0.0257 & 156.7 & 1.56\,$\times$\,10$^{-3}$ & 111 \\
18 18 0 1 - 17 17 1 1 & 348861.2642 & 0.0258 & 187.9 & 1.55\,$\times$\,10$^{-3}$ & 111 & Blended with CH$_3$OCHO and  \\
18 18 1 1 - 17 17 0 1 & 348861.2642 & 0.0258 & 187.9 & 1.55\,$\times$\,10$^{-3}$ & 37 & CH$_2$DOCHO + Absorption \\
22 15 8 1 - 21 14 7 1 & 350450.9471 & 0.0203 & 196.2 & 9.94\,$\times$\,10$^{-4}$ & 45 & Blended with CH$_3$CN \\
22 15 7 1 - 21 14 8 1 & 350450.9510 & 0.0203 & 196.2 & 9.94\,$\times$\,10$^{-4}$ & 135 & + Absorption \\
22 15 7 0 - 21 14 8 0 & 350502.5773 & 0.0204 & 165.6 & 1.02\,$\times$\,10$^{-3}$ & 45 & Blended with CH$_2$OHCHO \\
22 15 8 0 - 21 14 7 0 & 350502.6386 & 0.0204 & 165.6 & 1.02\,$\times$\,10$^{-3}$ & 135 \\
21 16 5 0 - 20 15 6 0 & 352806.0046 & 0.0170 & 164.3 & 1.19\,$\times$\,10$^{-3}$ & 129 & Blended with CH$_2$DCHO \\
21 16 6 0 - 20 15 5 0 & 352806.0050 & 0.0170 & 164.3 & 1.19\,$\times$\,10$^{-3}$ & 43 \\
21 16 6 1 - 20 15 5 1 & 352896.1707 & 0.0169 & 195.1 & 1.16\,$\times$\,10$^{-3}$ & 129 & Blended with HNCO \\
21 16 5 1 - 20 15 6 1 & 352896.1707 & 0.0169 & 195.1 & 1.16\,$\times$\,10$^{-3}$ & 43 & + Absorption \\
20 17 3 0 - 19 16 4 0 & 355022.2395 & 0.0174 & 164.0 & 1.37\,$\times$\,10$^{-3}$ & 41 & Blended with (CH$_2$OH)$_2$ \\
20 17 4 0 - 19 16 3 0 & 355022.2395 & 0.0174 & 164.0 & 1.37\,$\times$\,10$^{-3}$ & 123 & + Absorption \\
20 17 3 1 - 19 16 4 1 & 355248.4955 & 0.0174 & 195.0 & 1.35\,$\times$\,10$^{-3}$ & 123 & Blended with CH$_3$COCH$_3$ \\
20 17 4 1 - 19 16 3 1 & 355248.4955 & 0.0174 & 195.0 & 1.35\,$\times$\,10$^{-3}$ & 41 \\
19 18 2 0 - 18 17 1 0 & 357200.3403 & 0.0231 & 164.7 & 1.58\,$\times$\,10$^{-3}$ & 39 & Blended with CH$_3$CHO \\
19 18 1 0 - 18 17 2 0 & 357200.3403 & 0.0231 & 164.7 & 1.58\,$\times$\,10$^{-3}$ & 117 \\
23 15 9 1 - 22 14 8 1 & 359038.5618 & 0.0238 & 205.9 & 1.01\,$\times$\,10$^{-3}$ & 141 & Absorption\\
23 15 8 1 - 22 14 9 1 & 359038.5787 & 0.0238 & 205.9 & 1.01\,$\times$\,10$^{-3}$ & 47 \\
23 15 8 0 - 22 14 9 0 & 359134.3667 & 0.0239 & 175.3 & 1.04\,$\times$\,10$^{-3}$ & 141 & Blended with CH$_2$DOH \\
23 15 9 0 - 22 14 8 0 & 359134.6047 & 0.0239 & 175.3 & 1.04\,$\times$\,10$^{-3}$ & 47 & and CH$_3$CDO \\
22 16 6 0 - 21 15 7 0 & 361496.5122 & 0.0191 & 173.6 & 1.21\,$\times$\,10$^{-3}$ & 45 & Blended with (CH$_2$OH)$_2$ \\
22 16 7 0 - 21 15 6 0 & 361496.5152 & 0.0191 & 173.6 & 1.21\,$\times$\,10$^{-3}$ & 135 \\
\hline
\end{tabular}
\end{center}
\label{List_lines_excluded}
\end{table*}%

\section{Chemical modeling of 3-hydroxypropenal (HOCHCHCHO) and its isomers}
\label{appendix_chemistry}

HOCHCHCHO has a large number of isomers such as 2-propenoic acid (C$_2$H$_3$COOH), methyl glyoxal (CH$_3$COCHO), propanedial (HCOCH$_2$CHO), 2-hydroxypropenal (CH$_2$COHCHO), allenediol (HOCHCCHOH), vinyl formate (C$_2$H$_3$OCHO) and even cyclic compounds: $\beta$-propiolactone and oxiranecarboxaldehyde, with C$_2$H$_3$COOH the most stable isomer. 

Complex molecules, such as the ones with a C$_3$ skeleton, are mainly produced through either: i) grain reactions of unsaturated molecules (C$_3$, C$_3$H, C$_3$H$_2$, C$_3$O, H$_2$C$_3$O, ...)  initially produced in the gas phase then stuck on grains or ii) associations between radicals on grains \citep[e.g.,][]{Harada2010,Garrod2013,Garrod2017,Manigand2021}. In the first case, the unsaturated molecules have to react on grains with O or OH to form molecules such as HOCHCHCHO and its isomers. As OH is not very mobile on grains and easily reacts with H$_2$, only atomic oxygen through reactions with the C$_3$H$_x$O species can produce the isomers of C$_3$H$_4$O$_2$. 
However, O mainly reacts with the abundant radicals on the grains (HCO, CH$_2$OH, and CH$_3$O) and the production of C$_3$H$_4$O$_2$ isomers from the reactions of O is limited. In fact, only the isomer C$_2$H$_3$COOH is produced by this type of reaction in our network, with a rather low production. 
 
In contrast to the O + C$_3$H$_x$O reactions, the reactions with HCO are much more efficient because HCO is abundant and relatively mobile on the grains \citep{Wakelam2017,Minissale2016}. So, the reactions HCO + C$_2$H$_{x=1,3}$O are efficient to produce (directly of after hydrogenation) the isomers of C$_3$H$_4$O$_2$ if the C$_2$H$_{x=1,3}$O radicals are abundant. The C$_2$H$_{x=1}$O (HCCO) is not very abundant on grains, according to our network, and thus it cannot be an important source of C$_3$H$_4$O$_2$. However, C$_2$H$_3$O radicals are quite abundant on grains, thereby leading to the formation of C$_3$H$_4$O$_2$. 
 
Among the various C$_2$H$_3$O radicals, CH$_3$CO is the most abundant one. It is produced through addition of H on H$_2$CCO, but also through H atom abstraction of CH$_3$CHO and through the CH$_3$ + CO reaction on grains. Besides CH$_3$CO, there are three other C$_2$H$_3$O radicals: CH$_2$CHO, CH$_2$COH, and HCCHOH; CH$_2$CHO is produced through the reaction H$_2$CCO + H and through H atom abstraction of CH$_3$CHO. Furthermore, CH$_2$COH forms through addition of H on H$_2$CCO and H atom abstraction of C$_2$H$_3$OH, and HCCHOH is produced by a H atom abstraction of C$_2$H$_3$OH.

The isomers of HOCHCHCHO are then formed according to the following reactions:
CH$_3$COCHO (methyl glyoxal) is produced via the reaction HCO + CH$_3$CO.
HCOCH$_2$CHO (propanedial) is formed with the reaction between HCO and CH$_2$CHO which also produces (to a minor extent) C$_2$H$_3$OCHO (vinyl formate) because the single electron of CH$_2$CHO is partly delocalized on oxygen (H$_2$C$^\bullet$$-$CH$=$O $\longleftrightarrow$ H$_2$C$=$CH$-$O$^\bullet$); CH$_2$COHCHO (2-hydroxypropenal) results from the reactions between HCO and CH$_2$COH.
The newly detected species, HOCHCHCHO (3-hydroxypropenal), can be produced by the reaction HCO + CHCHOH, but the abundance of CHCHOH is low. 

One important point is that methyl glyoxal/2-hydroxypropenal and 3-hydroxypropenal/propanedial are tautomeric couples. They are interconvertible by chemical reaction, in this case via a transfer of hydrogen atom (keto-enolic equilibrium). 
The reactions producing the C$_3$H$_4$O$_2$ isomers are exothermic. For example HCO + CH$_2$CHO $\rightarrow$ HCOCH$_2$CHO is exothermic by 306 kJ/mol. In this case, HCOCH$_2$CHO is produced above the barrier to isomerization to HOCHCHCHO, located 72 kJ/mol below the HCO + CH$_2$CHO entrance energy. The HCOCH$_2$CHO with 306 kJ/mol of internal energy will relax through interaction with the ice. The typical timescale for isomeric conversion between HCOCH$_2$CHO and HOCHCHCHO is shorter than for relaxation. Thus, as relaxation occurs, isomeric conversion leads to equilibrated isomeric abundances at each internal energy. The final balance is determined at or near the effective barrier to isomerization, which corresponds to the energy of the transition state \citep{Herbst2000}, favoring HOCHCHCHO in that case. 
In our network, we consider that the ratio between the isomeric forms is then approximated by the ratio of the rovibrational densities of states of the isomers at the barrier to isomerization. For some reactions, the C$_3$H$_4$O$_2$ isomers are produced above dissociation limit and then some dissociate. The branching ratios towards dissociation are roughly estimated according to the excess energy above the dissociation limit and play a relatively small role in C$_3$H$_4$O$_2$ chemistry.

\begin{figure}[h!]
\begin{center}
\includegraphics[width=1\hsize]{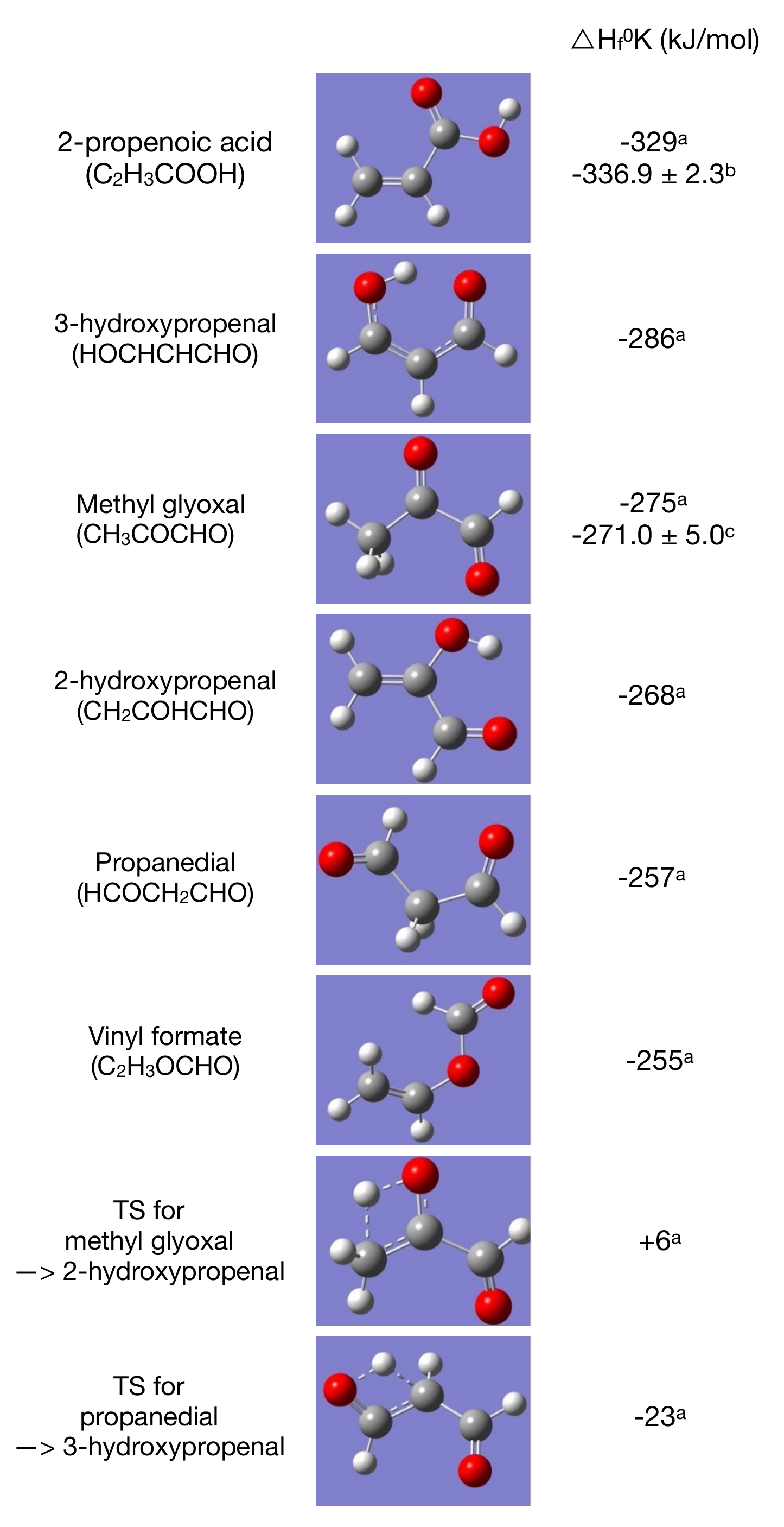}
\caption{Representation of the isomers of HOCHCHCHO included in the chemical modeling with their formation energies and the transition states (TS) between the tautomeric forms. The values with $^{\rm a}$ are obtained through M06-2X/AVTZ calculations using C$_2$H$_4$ + CO$_2$ as a reference, while the ones with $^{\rm b}$ and $^{\rm c}$ refer to \citet{Guthrie1978} and \citet{Moulds1938}, respectively.}
\label{fig_isomers}
\end{center}
\end{figure}

By considering the different reactions producing C$_3$H$_4$O$_2$, we introduced six of the isomers in our network: 3-hydroxypropenal (HOCHCHCHO), 2-propenoic acid (C$_2$H$_3$COOH), methyl glyoxal (CH$_3$COCHO), propanedial (HCOCH$_2$CHO), 2-hydroxypropenal (CH$_2$COHCHO), and vinyl formate (C$_2$H$_3$OCHO). These species are presented in Figure \ref{fig_isomers} with their respective formation energies and the transition states between the tautomeric forms. The considered reactions can be found in Table \ref{List_reactions}. 

\onecolumn
\begin{longtable}{@{}l@{ }l@{ }l@{~~}l@{~~}l@{~~}l@{}}
\caption{\label{List_reactions} Reactions involved in the formation of the C$_3$H$_4$O$_2$ isomers.} \\
\hline
\hline
Reaction & & $\Delta E$ & Branching & $\gamma$ (K) & Comments \\
                & & (kJ/mol)   & ratio           & (imaginary & \\
                & &                 &                   & frequency) & \\
\hline
\endfirsthead
\caption{continued.}\\
\hline
\hline
Reaction & & $\Delta E$ & Branching & $\gamma$ (K) & Comments \\
                & & (kJ/mol)   & ratio           & (imaginary & \\
                & &                 &                   & frequency) & \\
\hline
\endhead
\\
\multicolumn{6}{c}{C$_3$H$_4$O$_2$ production} \\
\\
\hline
HCO + CH$_3$CO & $\rightarrow$ CH$_3$COCHO & -300 & 0.2 & 0 & The TS for CH$_3$COCHO $\rightarrow$ CH$_3$CHO + CO is  \\
& $\rightarrow$ CH$_2$COHCHO &-292  & 0.05 & 0 & located 337 kJ/mol above the CH$_3$COCHO level \\
& $\rightarrow$ CH$_3$CHO + CO & -304 & 0.75 & 0 & so CH$_3$COCHO cannot dissociate. The TS for \\
& $\rightarrow$ H$_2$CCO + H$_2$CO & -196 & 0 & & CH$_3$COCHO $\rightarrow$ CH$_2$COHCHO is located 281\\
& $\rightarrow$ CH$_4$ + CO + CO &  -322 & 0 & &   kJ/mol above the CH$_3$COCHO level so CH$_3$COCHO \\
& $\rightarrow$ c-C$_2$H$_4$O + CO & -204 & 0 &  &   can isomerize. However, CH$_3$COCHO should \\
&   &  & &  &   be favorized. The TS for CH$_3$CHO $\rightarrow$ CH$_4$ + CO \\
&  &  & &  &  is located 354 kJ/mol above the CH$_3$CHO energy,\\
&   &  & &  &  so the CH$_3$CHO does not dissociate.\\
\hline
HCO + CH$_2$CHO & $\rightarrow$ HCOCH$_2$CHO & -306 & 0.05 & 0 & The TS for HCOCH$_2$CHO $\rightarrow$ HOCHCHCHO is  \\
& $\rightarrow$ HOCHCHCHO & -335 & 0.2 & 0 & located +234 kJ/mol above the HOCH$_2$CHO level \\
& $\rightarrow$ C$_2$H$_3$OCHO & -303 & 0.02 & 0 & and the TS for HCOCH$_2$CHO $\rightarrow$ CH$_3$CHO + CO\\
& $\rightarrow$ CH$_3$CHO + CO & -328 & 0.73 & 0 & is located 301 kJ/mol above the HOCH$_2$CHO level \\
& $\rightarrow$ H$_2$CCO + H$_2$CO & -220 & 0 &  & so HOCH$_2$CHO can isomerize but mostly not \\
& $\rightarrow$ CH$_4$ + CO + CO & -347 & 0 &  & dissociate. The lonely electron of CH$_2$CHO is  \\
& & & & & localized on CH$_2$ (80\%) site with a small but non  \\
& & & & & negligible contribution to oxygen (20\%). So, some  \\
& & & & & C$_2$H$_3$OCHO should be produced. \\
\hline
HCO + CH$_2$COH & $\rightarrow$ CH$_2$COHCHO & -408 & 0.05 & 0 & The TS for CH$_2$COHCHO $\rightarrow$ CH$_3$COCHO is \\
& $\rightarrow$ CH$_3$COCHO & -416 & 0.2 & 0 & located 242 kJ/mol above the CH$_2$COHCHO level  \\
& $\rightarrow$ CH$_3$CHO + CO & -481 & 0.75 & 0 & so CH$_2$COHCHO can isomerize. The TS for \\
& $\rightarrow$ H$_2$CCO + H$_2$CO & -312 & 0 & & CH$_3$COCHO $\rightarrow$ CH$_3$CHO + CO is  located 337  \\ 
& $\rightarrow$ CH$_4$ + CO + CO & -456 & 0 & &  kJ/mol above the CH$_3$COCHO level, so  CH$_3$COCHO \\
& & & & & can dissociate. \\
\hline
HCO + CHCHOH & $\rightarrow$ HOCHCHCHO & -444 & 0.2 & 0 & The TS for HOCHCHCHO $\rightarrow$ HCOCH$_2$CHO is  \\
& $\rightarrow$ HCOCH2CHO & -415 & 0.05 & 0 & located 263 kJ/mol above the HOCHCHCHO level \\
& $\rightarrow$ C$_2$H$_3$OH + CO & -396 & 0.75 & 0 & so HOCHCHCHO can isomerize but is favored. \\
& $\rightarrow$ HCCOH + H$_2$CO & -190 & 0 & & The TS for HCOCH$_2$CHO $\rightarrow$ CH$_3$CHO + CO is  \\
& $\rightarrow$ CH$_4$ + CO + CO & -456 & 0 & & located 301 kJ/mol above the HOCH$_2$CHO level  \\
& & & & & so HOCH$_2$CHO can dissociate. \\
\hline
\\
\multicolumn{6}{c}{C$_3$H$_4$O$_2$ consumption} \\
\\
\hline
H + C$_2$H$_3$COOH & $\rightarrow$ CH$_3$CHCOOH & -176 & 1 & 1000(800i) & Barrier and imaginary frequency are guessed by \\
& & & & & comparison with similar reactions. \\
\hline
H + CH$_3$COCHO & $\rightarrow$ CH$_3$CO + CO + H$_2$ & -66 & 1 & 3400(1900i) & The barriers and imaginary frequencies are  \\
& $\rightarrow$ CH$_3$COCH$_2$O & -89 & 0 & 2200(900i) & calculated at M06-2X/AVTZ level but for  the moment \\
& $\rightarrow$ CH$_3$COHCHO &  -201 & 0 & 3900(1200i) & we consider only one type of products to take into \\
& $\rightarrow$ CH$_3$COCHOH & -203 & 0 & 4150(1300i) &   account the consumption without introducing  all the \\
& & & & & C$_3$H$_5$O$_2$ isomers. \\
\hline
H + HOCHCHCHO &  $\rightarrow$ HOCH$_2$CHCHO & -107 & 0 & 2900(840i) & The barriers and imaginary frequencies are  \\
&  $\rightarrow$ HOCHCHCH$_2$O & -19 & 0 & 4100(900i) & calculated at M06-2X/AVTZ level but for  the moment \\
&  $\rightarrow$ HOCHCHCHOH & -111 & 0 & 4000(1300i) & we consider only one type of products to take into \\
&  $\rightarrow$ HOCHCHCO + H$_2$ & -20 & 0 & 3100(1600i) & account the consumption without introducing  all the \\
&  $\rightarrow$ CH$_3$CHO + HCO & -88 & 1 & 3100(1600i) & C$_3$H$_5$O$_2$ isomers. \\
\hline
H + HCOCH$_2$CHO & $\rightarrow$ CH$_3$CHO + HCO & -82 & 1 & 3000(1600i) & Barrier and imaginary frequency are guessed by \\
& & & & & comparison with similar reactions. \\
\hline
H + CH$_2$COHCHO & $\rightarrow$ CH$_3$COHCHO & -209 & 0 & 1100(500i) & The barriers and imaginary frequencies are  \\
& $\rightarrow$ CH$_3$CHO + HCO & -72 & 1 & 1100(500i) & calculated at M06-2X/AVTZ level but for  the moment \\
& & & & & we consider only one type of products to take into \\
& & & & & account the consumption without introducing  all the \\
& & & & & C$_3$H$_5$O$_2$ isomers. \\
\hline
H + C$_2$H$_3$COCHO & $\rightarrow$ CH$_3$CHOCHO & -150 & 0 & 800(630i) & The barriers and imaginary frequencies are  \\
& $\rightarrow$ CH$_3$CHO + HCO & -85 & 1 & 800(630i) & calculated at M06-2X/AVTZ level but for the moment \\
& & & & & we consider only one type of products to take into \\
\hline
& & & & & account the consumption without introducing  all the \\
& & & & & C$_3$H$_5$O$_2$ isomers. \\
\hline
\\
\multicolumn{6}{c}{C$_2$H$_3$O production} \\
\\
\hline
H + H$_2$CCO & $\rightarrow$ CH$_3$CO & -165 & 0.1 & 2300(680i) & The barriers and imaginary frequencies are  \\
& $\rightarrow$ CH$_3$ + CO & -120 & 0.3 & 2300(680i) & calculated at M06-2X/AVTZ level. The TS for  \\
& $\rightarrow$ CH$_2$CHO & -141 & 0.3 & 4030(930i) & CH$_3$CO $\rightarrow$ CH$_3$ + CO is located 69 kJ/mol above   \\
& $\rightarrow$ CH$_2$COH & -49 & 0.3 & 7000(1600i) & the CH$_3$CO energy, so a large part of CH$_3$CO \\
& & & & & dissociates. The TS for CH$_2$CHO $\rightarrow$ CH$_3$CO is \\
& & & & & located 166 kJ/mol above the CH$_2$CHO energy and  \\
& & & & & then CH$_2$CHO cannot isomerize. The TS for  \\
& & & & & CH$_2$COH $\rightarrow$ CH$_2$CHO is located 139 kJ/mol above \\
& & & & & the CH$_2$COH energy, so CH$_2$COH cannot  \\
& & & & & isomerize. See also \citet{Senosiain2006} and \\
& & & & & \citet{Michael1979}. \\
\hline
H + HCCOH & $\rightarrow$ CH$_2$COH & -188 & 0.1 & 1840(680i) & The barriers and imaginary frequencies are  \\
& $\rightarrow$ CHCHOH & -171 & 0.2 & 1920(770i) & calculated at M06-2X/AVTZ level. The TS for  \\
& $\rightarrow$ CH$_2$CHO & -281 & 0.7 & 1840(680i) & CH$_2$COH $\rightarrow$ CH$_2$CHO is located 139 kJ/mol above \\
& $\rightarrow$ H$_2$ + HCCO & -129 & 0 & & the CH$_2$COH energy, so most of the CH$_2$COH  \\
& & & & & isomerize into CH$_2$CHO. The TS for CHCHOH \\ 
& & & & & $\rightarrow$ CH$_2$CHO  is located 146 kJ/mol above the \\
& & & & & CHCHOH energy, and the TS for CHCHOH $\rightarrow$  \\
& & & & & CH$_2$COH is located 199 kJ/mol above the  \\
& & & & & CHCHOH  energy, so some CHCHOH isomerizes   \\
& & & & & into CH$_2$CHO but not into CH$_2$COH. \\
\hline
H + CH$_3$CHO & $\rightarrow$ C$_2$H$_5$O & -69 & 0.25 & 3300(910i) & The barriers and imaginary frequencies are  \\
& $\rightarrow$ CH$_3$CHOH & -110 & 0.25 & 5040(1360i) & calculated at M06-2X/AVTZ level. \\
& $\rightarrow$ CH$_3$CO + H$_2$ & -61 & 0.25 & 2400(1660i) & See also \citet{Hippler2002} and  \\
& $\rightarrow$ CH$_2$CHO + H$_2$ & -38 & 0.25 & 4900(1700i) & \citet{Whytock1976}. \\
\hline
CH$_2$ + HCO &  $\rightarrow$ CO + CH$_3$ & -383 & 0.7 & 0 & The TS for CH$_2$CHO $\rightarrow$ CH$_3$CO is located \\
&  $\rightarrow$ CH$_2$CHO & -405 & 0.1 & 0 & 166 kJ/mol above the CH$_2$CHO level, \\
&  $\rightarrow$ c-CH$_2$(O)CH & -266 & 0 & 0 & so -239 kJ/mol below the CH$_2$ + HCO energy. \\
&  $\rightarrow$ CH$_3$CO & -429 & 0.1 & 0 & \\
&  $\rightarrow$ H$_2$CCO + H & -264 & 0.1 & 0 & \\
\hline
CH$_3$ + CO & $\rightarrow$ CH$_3$CO & -46 & 1 & 3500(420i) & The barrier and imaginary frequency is \\
& & & & & calculated at M06-2X/AVTZ level taking into  \\
& & & & & account the previous results from \\
& & & & &  \citet{Baulch2005}, \citet{Anastasi1982},\\
& & & & &  \citet{Senosiain2006}, and \citet{Ding2001}. \\
\hline
O + C$_2$H$_3$ & $\rightarrow$ CH$_2$CHO & -536 & 0.2 & 0 & Deduced from \citet{Jung2015}. \\
& $\rightarrow$ CH$_3$CO & -559 & 0.2 & 0 \\
& $\rightarrow$ H$_2$CCO + H & -380 & 0.2 & 0 \\
& $\rightarrow$ CH$_3$ + CO & -512 & 0.4 & 0 \\
& $\rightarrow$ C$_2$H$_2$ + OH & -275 & 0 \\
\hline
\end{longtable}

\end{document}